\documentclass[aip,graphicx]{revtex4-1}
\usepackage{graphicx}
\usepackage{amsmath,amssymb}
\usepackage{color}

\begin{document}

\title{Self-propelled motion of a fluid droplet under chemical reaction}

\author{S. Yabunaka}
\email[]{yabunaka@scphys.kyoto-u.ac.jp}
\address{Department of Physics, Kyoto University, Kyoto 606-8502, Japan}
\author{T. Ohta}
\email[]{takao@scphys.kyoto-u.ac.jp}
\address{Department of Physics, Kyoto University, Kyoto 606-8502, Japan}
\author{N. Yoshinaga}
\email[]{yoshinaga@wpi-aimr.tohoku.ac.jp}
\address{WPI-AIMR, Tohoku University, Sendai, 980-8577, Japan}


%


\date{\today}

\begin{abstract}
We study self-propelled dynamics of a droplet due to a Marangoni
effect and chemical reactions in a binary fluid with a dilute third
 component of chemical product which affects the interfacial energy of a
 droplet. 
The equation for the  migration velocity of the center
of mass of a droplet is derived in the limit of  an infinitesimally
thin interface. 
We found that there is a bifurcation from a motionless state to a propagating state
of droplet by changing the strength of the Marangoni effect.
\end{abstract}


\maketitle

\section{Introduction}

Self-propelled motion of particles has attracted much attention recently
from the viewpoint of non-linear physics
far from equilibrium.
There are several experiments of self-propulsion of droplets in fluids
\cite{Nagai,Toyota1,Ban, Thutupalli}. 
It has been shown that the Belousov-Zhabotinsky reaction
composed in a fluid droplet triggers a spontaneous motion of a droplet \cite{Kitahata}.
Computer simulations of convective droplet motion \cite{Yeomans} and
nano-dimer motors \cite{Kapral1, Kapral2} driven by chemical reactions
have also been carried out. There are theoretical studies of droplet
motion due to an interfacial tension gradient along the droplet surface
\cite{Kitahata, Levan,Ryazantsev}. 
However, these theories are concerned only with the steady velocity of a droplet.
As a related theoretical study, the mesoscopic description of the thermo-capillary effect has been 
formulated \cite{Jasnow}. A transition between a motionless and migrating droplet driven by chemical reactions has been studied in 
a system where a droplet is on a solid substrate \cite{Baer}.

It should be noted that self-propelled motion of particles has been
investigated in a different field of physics. It has been known that a
pulse or a domain in excitable reaction diffusion systems exhibits a
bifurcation from a motionless state to a propagation state by changing
the system parameters \cite{Krischer,Or-Guil}. 
A reaction-diffusion system is represented by a set of  nonlinear partial differential
equations, that is often investigated by numerical simulations due to the
limitation of analytical calculations.
Nevertheless, the theory of domain dynamics in the vicinity of this
drift bifurcation has been developed, e.g., for  the interaction between domains
\cite{Ohta2001,Ei,Nishiura05} and for deformations of domain \cite{OOS,SHO, OhtaOhkuma}.

The purpose of the present paper is to extend the previous studies in
reaction-diffusion systems to the droplet motion in chemically
reacting fluids. We introduce a model system of binary fluids where a
chemical reaction takes place inside a droplet.  The chemical component
produced diffuses away from the droplet and influences the interfacial
energy.  The long range hydrodynamic effects are treated with a
Stokes approximation supposing that the relaxation of the fluid velocity field is much faster than that of the concentrations  and that the Reynold number is sufficiently small in the system considered. We will show that there is a drift bifurcation at certain threshold of the Marangoni strength as in the reaction-diffusion systems mentioned above. The time-evolution equation of the center of mass of droplet is derived near the drift bifurcation by taking into consideration of the hydrodynamic effects.

In the next section (section \ref{model}), we describe our model system and the interface dynamics. The equation of motion for the center of mass is derived in section \ref{equation.motion}. Discussion is given in section \ref{discussion}. The force acting on the droplet interface is formulated in Appendix \ref{forces}. Some of the details in the derivation of the velocity of the center of mass are given in Appendix \ref{Derivation-of-velocity}. The formulas used in the evaluation of the coefficients in the time-evolution equation for a droplet are summarized in Appendix \ref{the-evaluation-of}.  The convective effect of the third chemical component is estimated in Appendix \ref{convection}.

\section{Model and Interface dynamics}
\label{model}

We consider a fluid mixture where the free energy is given in terms of
the local concentration difference $\phi=\phi_{A}-\phi_{B}$ by 
\begin{eqnarray}
F\{\phi\}=
\int d\vec{r}
\left[
\frac{B(c)}{2}(\vec{\nabla}\phi)^{2}
+f_{\rm GL} (\phi)
+ f_0 (c)
\right],
\label{Free}
\end{eqnarray}
where $\phi_A (\phi_B)$ is the local concentration of the component A
(B) and $f_0(c)= c\ln c$.
 The coefficient $B>0$ is assumed to 
depend on $c$ as $B(c)=B_{0}+B_{1}c$ with $B_0$ and $B_1$ constants 
and 
$f_{\rm GL}(\phi)$ is a function
of $\phi$ such that phase separation takes place at  low temperatures.
Here we have assumed existence of a dilute third component whose concentration
is denoted by $c$. The logarithmic term ($ f_0 (c) = c \ln c$) arises from the
translational entropy of the dilute component. 
 The spatial variation of
$c$ is also assumed to be broad enough compared to that of $\phi$
which constitutes a sharp interface.

The time-evolution equation for $\phi$ is given by 
\begin{eqnarray}
\frac{\partial\phi}{\partial t}+
\vec{\nabla} \cdot\left(\vec{v}\phi\right)=\nabla^{2}\frac{\delta F}{\delta\phi},
\label{phi}
\end{eqnarray}
 where $\vec{v}$ is the local velocity whose equation is given by
eq. (\ref{stokes}) below. Hereafter we consider an isolated droplet
such that the concentration variation is $\phi(x)=\phi_{e}>0$ inside
the droplet and $\phi(x)=-\phi_e$ at the surrounding matrix. The equilibrium
value $\phi_{e}$ is determined by equating the rhs of eq. (\ref{phi}) to zero.
The dilute component $c$ is assumed to obey 
\begin{eqnarray}
\frac{\partial c}{\partial t}+\vec{\nabla}\cdot\left(\vec{v}c\right)=D\nabla^{2}c-\gamma (c-c_{\infty})+A\theta\left(R-\left|\vec{r}-\vec{r}_{G}\right|\right),
\label{composition}
\end{eqnarray}
where $\theta(x)$ is the step function such that $\theta(x)=0$
for $x\leq0$ and $\theta(x)=1$ for $x\geq0$. 
 The first term on the rhs of eq. (\ref{composition}) arises from
 $\vec{\nabla} \cdot [L(c)\vec{\nabla} \delta F/\delta c]$
with $L(c)=Dc$ where $D$ is positive constant. The $c$-dependence
of the Onsager coefficient $L$ is necessary for a dilute component \cite{Oono}.
The second term in eq. (\ref{composition}) indicates consumption of
$c$ with the rate $\gamma>0$ due to a chemical reaction and with $c=c_{\infty}$ for $|\vec{r}| \to \infty$ whereas
the last term represents production of $c$, which occurs  inside a droplet with radius $R$, whose center of mass is 
located at $\vec{r}_{G}$. In the most parts of the present paper, the coefficient $A$ is assumed to be positive and 
stands for the strength of the production. However the theory can also hold for $A<0$ with a slight modification.

The Stokes approximation is employed for the local velocity $\vec{v}$ and it takes the
form 
\begin{eqnarray}
0=-\vec{\nabla}p-\phi\vec{\nabla}\frac{\delta F}{\delta\phi}-c\vec{\nabla}\frac{\delta F}{\delta c}+\eta_{0}\nabla^{2}\vec{v} ,
\label{stokes}
\end{eqnarray}
 where $p$ is determined such that the velocity field satisfies the incompressibility
condition $\vec \nabla\cdot\vec{v}=0$. The viscosity $\eta_{0}$ is assumed, for simplicity, to be a
constant independent of $\phi$. The force arising from the first, second and third terms can be
written as 
\begin{eqnarray}
f^{\alpha}= - \nabla^{\alpha} p-\phi \nabla^{\alpha}\frac{\delta F}{\delta\phi}-c\nabla^{\alpha}\frac{\delta F}{\delta c}=-\nabla^{\alpha}p''+f_{\parallel}^{\alpha}+f_{\perp}^{\alpha}  ,
\label{f}
\end{eqnarray}
 where $p''$ has some additive terms to $p$, whose explicit form
is unnecessary for incompressible fluids since only the transverse
components of the velocity is relevant.  In Appendix \ref{forces}, we show that the normal and tangential
forces are given, respectively, by 
\begin{eqnarray}
f_{\parallel}^{\alpha}&=&-n^{\alpha}B(c)|\vec{\nabla}\phi|^{2}(\vec{\nabla}\cdot\vec{n})  ,
\label{f1}\\
f_{\perp}^{\alpha}&=&(\delta_{\alpha\beta}-n^{\alpha}n^{\beta})(\nabla^{\beta}B)|\vec{\nabla}\phi|^{2}  ,
\label{f2}
\end{eqnarray}
 where the unit vector $\vec{n}$ is directed to the outside of the droplet,
i.e., $\vec{n}=-\vec{\nabla}\phi/|\vec{\nabla}\phi|$. The repeated
indices imply the summation.  When we are concerned with the large scale
compared with the interface width (or the sharp interface limit), the
factor $|\vec{\nabla}\phi|^{2}$ is localized in the interface region. 
In this situation, 
the forces are localized on the interface at  $a$ which denotes a
location on the interface so that 
we may rewrite Eqs.  (\ref{f1}) and (\ref{f2}), respectively, as
\begin{eqnarray}
f_{\parallel}^{\alpha}(a)&=&-n^{\alpha}\sigma(a)(\vec{\nabla}\cdot\vec{n})  ,
\label{f3}\\
f_{\perp}^{\alpha}(a)&=&(\delta_{\alpha\beta}-n^{\alpha}n^{\beta})(\nabla^{\beta}\sigma)_I   ,
\label{f4}
\end{eqnarray}
The interfacial tension is defined by 
\begin{eqnarray}
\sigma(a)=\int dwB(c)\left(\frac{\partial\phi}{\partial w}\right)^{2}\approx B_{I}(a)\int dw\left(\frac{\partial\phi}{\partial w}\right)^{2}  ,
\label{tension}
\end{eqnarray}
where $w$ is the coordinate along the normal to the interface and 
$B_{I}$ is the value of $B$ at the interface. 
It should be noted that the  derivative in $\nabla^{\beta}\sigma$ is
not restricted to the two-dimensional space on the interface regarding $\sigma(a)$ as $\sigma(c(\vec{r}))$.
After taking the derivative in three dimensions, we may take the value on the interface.
This interpretation is consistent with Eq.  (\ref{f2}) in which $\nabla^{\beta}$ acts on the weak spatial variation of $c$.
The tangential component is automatically extracted by the projection $(\delta_{\alpha\beta}-n^{\alpha}n^{\beta})$.
Equations (\ref{f3}) and (\ref{f4})
are consistent with the boundary condition employed in 
hydrodynamics with multi-component fluids \cite{Anderson}.

Substituting Eq. (\ref{f}) into Eq. (\ref{stokes}) and using the
incompressibility condition, the local velocity of fluid is given
by 
\begin{eqnarray}
v^{\alpha}(\vec{r},t) & = & \int da'T^{\alpha\beta}\left(\vec{r},\vec{r}(a')\right)n^{\beta}(a')\sigma(a',t)K(a',t)\nonumber \\
 & + & \int da'T^{\alpha\beta}\left(\vec{r},\vec{r}(a')\right)[\delta_{\beta\gamma}-n^{\beta}(a')n^{\gamma}(a')](\nabla^{\gamma}\sigma)_{I}  ,
 \label{lv}
 \end{eqnarray}
 where 
  $da'$
is the infinitesimal area on the interface. The integral is taken
all over the interface. The Oseen tensor is given by 
\begin{eqnarray}
T^{\alpha\beta}\left(\vec{r},\vec{r}'\right)=\frac{1}{8\pi\eta_{0}s}\Big[\delta_{\alpha\beta}+\frac{s^{\alpha}s^{\beta}}{s^{2}}\Big]  ,
\end{eqnarray}
 with $\vec{s}=\vec{r}-\vec{r}'$. The mean curvature is defined by
$K=-\vec{\nabla}\cdot\vec{n}$. 

The right hand side in the time-evolution equation (\ref{phi}) for $\phi$  can be ignored when
the hydrodynamic effects are dominant  \cite{Kawasaki}. From the left hand side  of Eq. (\ref{phi}),
we note that the normal component $V(a,t)$ of the interface velocity
is given by
\begin{eqnarray}
V(a,t)=v^{\alpha}(\vec{r}(a),t)n^{\alpha}(a). \label{Va}
\end{eqnarray}
Substituting Eq. (\ref{lv}) into Eq. (\ref{Va}), we obtain 
 \begin{eqnarray}
V(a,t)=V_1+V_2  ,
\label{V1}
\end{eqnarray}
 where 
 \begin{eqnarray}
V_1=\int da'n^{\alpha}(a)T^{\alpha\beta}\left(\vec{r}(a),\vec{r}(a')\right)n^{\beta}(a')\sigma(a',t)K(a',t)  ,
\label{V2}
\end{eqnarray}
 and 
 \begin{eqnarray}
V_2=\int da'n^{\alpha}(a)T^{\alpha\beta}\left(\vec{r}(a),\vec{r}(a')\right)[\delta_{\beta\gamma}-n^{\beta}(a')n^{\gamma}(a')](\nabla^{\gamma}\sigma)_{I}  .
\label{V3}
\end{eqnarray}

The velocity $\vec{u}$ of the center of mass of an isolated droplet
can be obtained from $V(a,t)$. The geometrical consideration leads
to \cite{Kawasaki}
\begin{eqnarray}
u^{\alpha}=\frac{1}{\Omega}\int daV(a)R^{\alpha}(a),
\label{velocity}
\end{eqnarray}
 where $\Omega$ is the volume of the droplet and $\vec{R}(a)$ is
the position vector directed from the center of mass to the interface.
For a spherical droplet with radius $R$, we have $\Omega=4\pi R^{3}/3$
and $\vec{R}(a)=\vec{n} (a) R$.

In order to determine the migration velocity $\vec{u}$, we have to
evaluate the interfacial tension and its spatial derivative as  Eqs.  (\ref{V2}) and  (\ref{V3}), which may depend on the concentration $c$. 
In this way, we take into account the Marangoni effect. To this end, we assume that  the interfacial tension depends on $c_I$ as
\begin{eqnarray}
\sigma=\sigma_{0}+\sigma_{1}c_{I},
\label{sigma}
\end{eqnarray}
 where $\sigma_{0}$ and $\sigma_{1}$ are constants determined from the expression of $B=B_0+B_1c$. However, the explicit form of  $\sigma_{0}$ and $\sigma_{1}$ as a function of $B_0$ and $B_1$ are unnecessary in the argument below. Substituting
(\ref{sigma}) into (\ref{velocity}),
we obtain for a spherical droplet with $K=-2/R$
\begin{eqnarray}
u^{\alpha} =u_{1}^{\alpha} + u_{2}^{\alpha} ,
\label{u}
\end{eqnarray}
where
\begin{eqnarray}
u_{1}^{\alpha} & =- & \frac{2\sigma_{1}}{\Omega}\int dan^{\alpha}(a)\int da'n^{\beta}(a)T^{\beta\gamma}(\vec{r}(a),\vec{r}(a'))n^{\gamma}(a')c_{I}(a')  ,
 \label{upara}\\
u_{2}^{\alpha} & =&\frac{\sigma_1R}{\Omega}\int dada'n^{\alpha}(a)n^{\delta}(a)T^{\delta\beta}\left(\vec{r}(a),\vec{r}(a')\right) \nonumber \\
&\times&(\delta_{\beta \gamma} - n^{\beta}(a') n^{\gamma} (a'))\nabla^{\gamma}c(a')  .
\label{uperp}
 \end{eqnarray}
Equations (\ref{upara}) and (\ref{uperp}) are derived
in  Appendix \ref{Derivation-of-velocity} as
\begin{eqnarray}
u_{1}^{\alpha} &=& -\frac{8\sigma_{1}R}{15\Omega \eta_0}\int da'n^{\alpha}(a')c_{I}(a')  ,
 \label{u1} \\
u_{2}^{\alpha} & = & 
  \frac{\sigma_{1}R^{2}}{5\Omega\eta_0}\int da'\left(\delta_{\alpha\delta}-n^{\alpha}(a')n^{\delta}(a')\right)\left(\nabla^{\delta}c\right)_I  .
   \label{u2}
   \end{eqnarray}
In the next section, we will derive the time-evolution equation for  $\vec{u}$  from Eq. (\ref{u}) with (\ref{upara}) and (\ref{uperp}) by solving  Eq. (\ref{composition}) for the third component $c$.


It is remarked that,
when $c(\vec{r})$ is set as $c= c_0 +
c_1 z$ instead of solving Eq. (\ref{composition}), we obtain from Eq. (\ref{u}) with
(\ref{upara}) and (\ref{uperp}) the stationary migration velocity $u = -2 \sigma_1 c_1 R/(15 \eta_0)$ which agrees with the
known result obtained by the conventional theory of the Marangoni effect \cite{Young}.

\section{Equation of motion for a droplet}
\label{equation.motion}

In this section, we derive the equation of motion for a droplet. Since the major hydrodynamic effects have been taken into account as in Eqs. (\ref{V1}),  (\ref{V2}) and  (\ref{V3}), we ignore the convective term $\vec{\nabla}\cdot\left(\vec{v}c\right)$ in Eq. (\ref{composition}). 
We will show in Appendix \ref{convection} and in section \ref{discussion} that 
this term causes a shift of the bifurcation threshold but  is not expected to
change the bifurcation behavior essentially. 

The configuration of the component $c$ around a droplet can be obtained by solving the following equation
\begin{eqnarray}
\frac{\partial c}{\partial t}=D\nabla^{2}c-\gamma (c-c_{\infty})+A\theta\left(R-\left|\vec{r}-\vec{r}_{G}\right|\right)  .
\label{EOMofC}
\end{eqnarray}
Hereafter, we consider the case of $A>0$ that the component $c$ is produced inside the droplet, diffuses away, and vanishes at $|\vec{r}| \to \infty$ i.e., $c_{\infty}=0$. The method can also be applied for $A<0$ with the boundary condition $c=c_{\infty}\ne 0$ for $|\vec{r}| \to \infty$.
In terms of the Fourier transform, Eq. (\ref{EOMofC}) can be written
as
\begin{eqnarray}
\frac{\partial c_{\vec{q}}}{\partial t} & = & -D\left(q^{2}+\beta^{2}\right)c_{\vec{q}}+H_{\vec{q}}  ,
\label{eqcq}
 \end{eqnarray}
where
\begin{eqnarray}
\beta=\left(\frac{\gamma}{D}\right)^{\frac{1}{2}},\\
H_{\vec{q}}=AS_{q}e^{i\vec{q}\cdot\vec{r}_{G}}  ,
\end{eqnarray}
with the form factor of a sphere
\begin{eqnarray}
S_{q} & = & \int d^{3}r\exp\left(i\vec{q}\cdot\vec{r}\right)\theta(\left|\vec{r}\right|-R)\\
 & = & 4\pi\frac{\sin(qR)-qR\cos(qR)}{q^{3}}  .
 \end{eqnarray}
The Fourier component $c_{\vec{q}}$ has been defined as 
\begin{eqnarray}
c_{\vec{q}}=\int d^{3}rc(\vec{r})e^{i\vec{q}\cdot\vec{r}}  .
\end{eqnarray}

By assuming the relaxation of the composition $c$ is sufficiently
rapid compared to the motion of interface, we solve Eq. (\ref{eqcq})
by means of an expansion in terms of the time derivative.
\begin{eqnarray}
c_{\vec{q}}&=&G_{q}H_{\vec{q}}-G_{q}^{2}\frac{\partial H_{\vec{q}}}{\partial t}+G_{q}^{3}\frac{\partial^{2}H_{\vec{q}}}{\partial t^{2}}-G_{q}^{4}\frac{\partial^{3}H_{\vec{q}}}{\partial t^{3}}+.....  \nonumber \\
&=&c_{\vec{q}}^{(0)}+c_{\vec{q}}^{(1)}+c_{\vec{q}}^{(2)}+c_{\vec{q}}^{(3)}+.....  ,
\label{c}
\end{eqnarray}
where we have defined
\begin{eqnarray}
G_{q}=\frac{1}{D\left(q^{2}+\beta^{2}\right)}.
\end{eqnarray}
The short time expansion (\ref{c}) is justified in the vicinity of  the supercritical drift bifurcation where the velocity of a droplet $u=|\vec{u}|$ is arbitrarily small. That is, 
the smallness parameter of this expansion is given by
 \begin{eqnarray}
\varepsilon=\frac{u}{D\beta} << 1, 
\label{u23}
\end{eqnarray} 
where the denominator is the characteristic time of $c$.
After the inverse Fourier transform, the composition $c_I$ at the interface is given by
\begin{eqnarray}
c_I=c_{I}^{(0)}(\vec{r}_{G}+\vec{s})+c_{I}^{(1)}(\vec{r}_{G}+\vec{s})+c_{I}^{(2)}(\vec{r}_{G}+\vec{s}) +c_{I}^{(3)}(\vec{r}_{G}+\vec{s})    ,
\end{eqnarray}
where
\begin{eqnarray}
c_{I}^{(0)}(\vec{r}_{G}+\vec{s})
&=&
A\int_{\vec{q}}G_{q}S_{q}e^{i\vec{q}\cdot\vec{r}_{G}}e^{-i\vec{q}\cdot(\vec{r}_{G}+\vec{s})}
=
A\int_{\vec{q}}G_{q}S_{q}e^{-i\vec{q}\cdot\vec{s}}    ,
\label{c0} \\
c_{I}^{(1)}(\vec{r}_{G}+\vec{s})&=&-A\int_{\vec{q}}(i\vec{q}\cdot\vec{u})G_{q}^{2}S_{q}e^{-i\vec{q}\cdot\vec{s}}=u^{\alpha}\frac{\partial}{\partial s^{\alpha}}Q_{2}(s)   ,
\label{c1}\\
c_{I}^{(2)}(\vec{r}_{G}+\vec{s}) & = &
A\int_{\vec{q}}(i\vec{q}\cdot\dot{\vec{u}})G_{q}^{3}S_{q}e^{-i\vec{q}\cdot\vec{s}}+A\int_{\vec{q}}(i\vec{q}\cdot\vec{u})^{2}G_{q}^{3}S_{q}e^{-i\vec{q}\cdot\vec{s}}\nonumber 
\\
 & = & -\dot{u}^{\alpha}\frac{\partial}{\partial s^{\alpha}}Q_{3}(s)+u^{\alpha}u^{\beta}\frac{\partial}{\partial s^{\alpha}}\frac{\partial}{\partial s^{\beta}}Q_{3}(s)   ,
 \label{c2}\\
c_{I}^{(3)}(\vec{r}_{G}+\vec{s}) & = & -A\int_{q}(i\vec{q}\cdot\vec{u})^{3}G_{q}^{4}S_{q}e^{-i\vec{q}\cdot\vec{s}}\nonumber \\
 & = & u^{\alpha}u^{\beta}u^{\gamma}\frac{\partial}{\partial s^{\alpha}}\frac{\partial}{\partial s^{\beta}}\frac{\partial}{\partial s^{\gamma}}Q_{4}(s)   .
 \label{c3}
 \end{eqnarray}
 The terms with the higher order time derivatives have been ignored.
The migration velocity is given by
\begin{eqnarray}
\vec{u}=\frac{d\vec{r}_G}{dt}.
\end{eqnarray}
We have defined $Q_{n}(s)$ by
\begin{eqnarray}
Q_{n}(s)=A\int_{\vec{q}}G_{q}^{n}S_{q}e^{-i\vec{q}\cdot\vec{s}}.
\end{eqnarray}

\begin{figure}[t]
\includegraphics[scale=0.3]{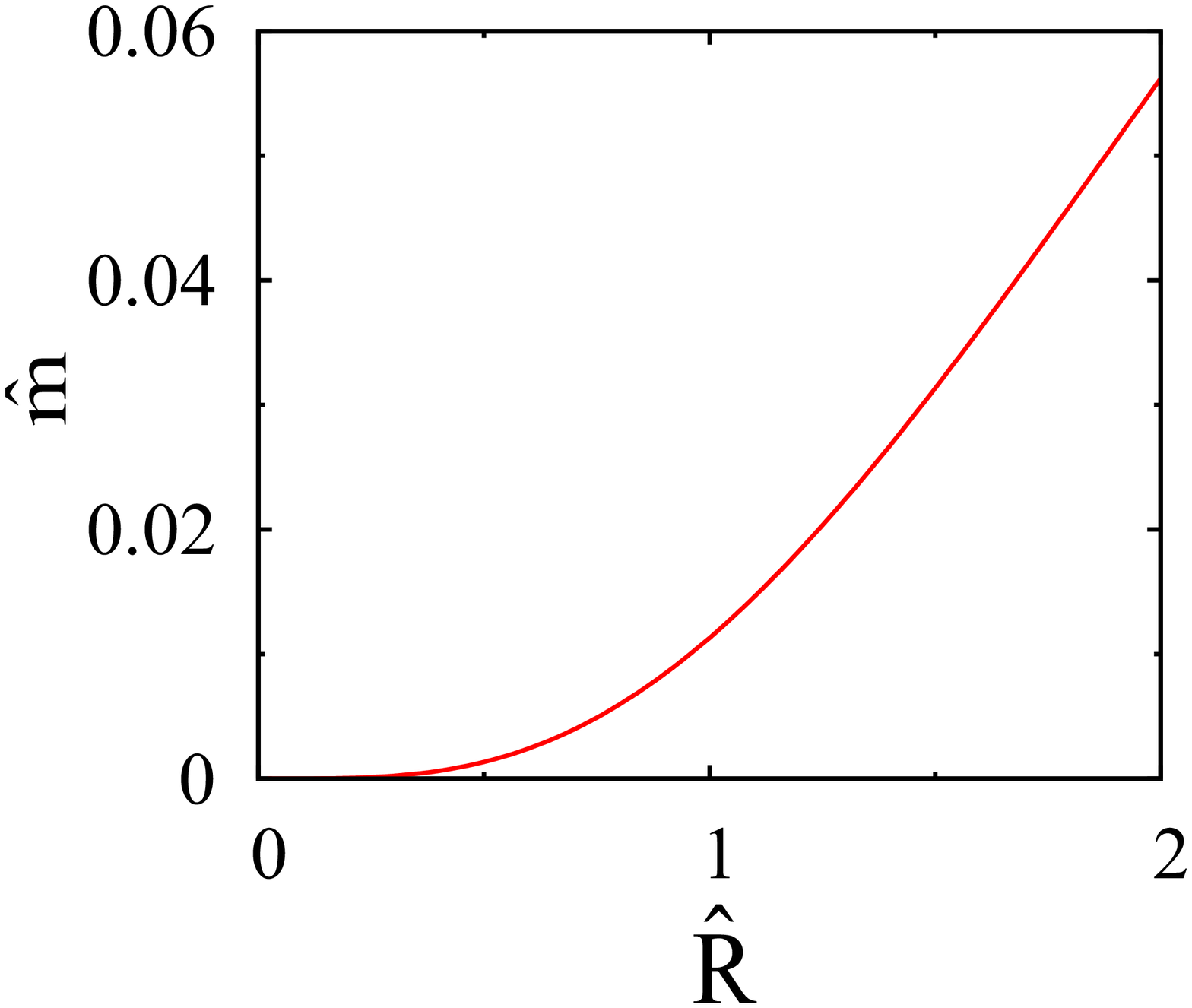}
\caption{The scaled coefficient $\hat{m}$  as a function of $\hat{R}$.}
\label{Flo:rescaled-1} 
\end{figure}

\begin{figure}[t]
\includegraphics[scale=0.3]{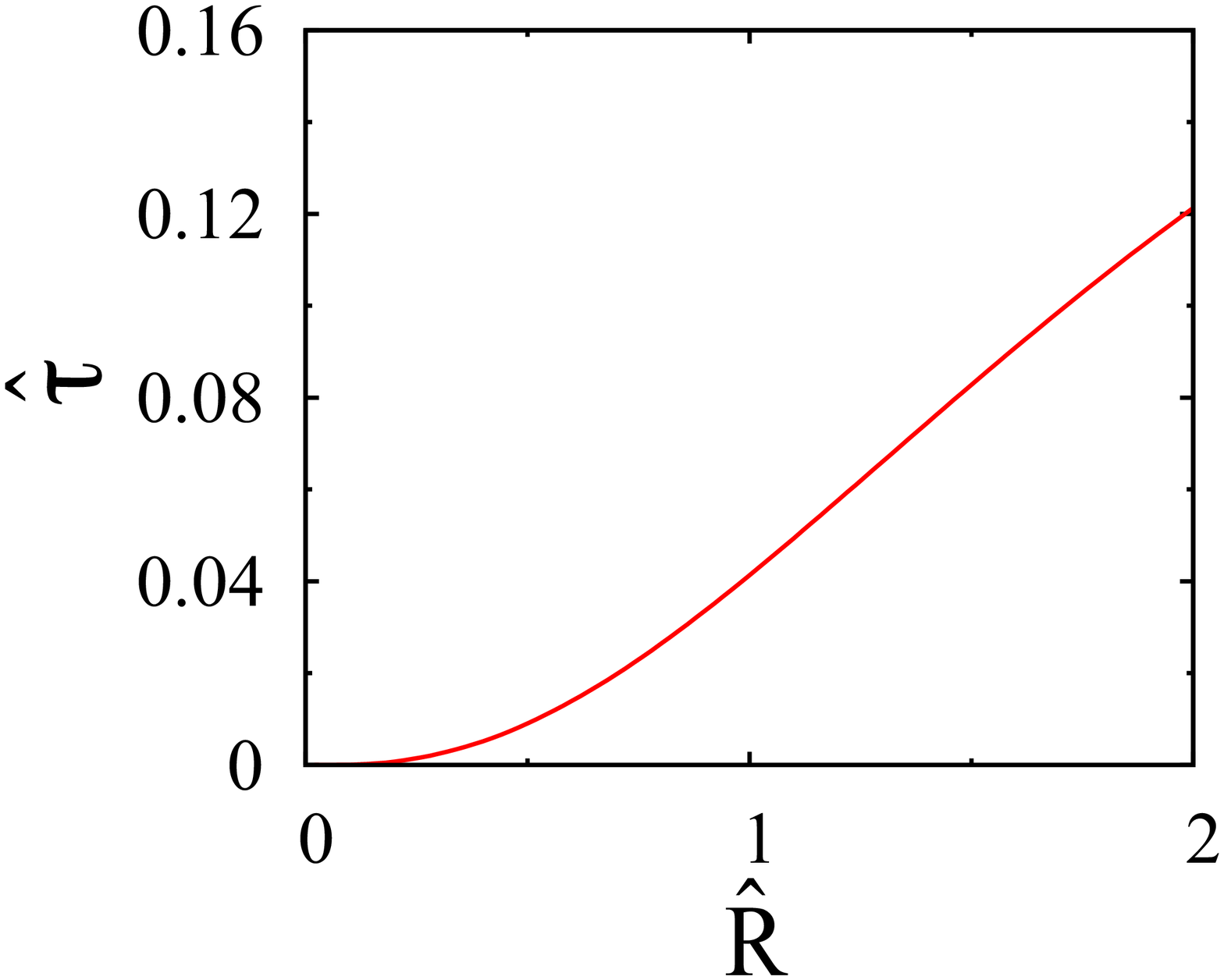}
\caption{The scaled coefficient $\hat{\tau}$ as a function of $\hat{R}$.}
\label{Flo:rescaled-2} 
\end{figure}

\begin{figure}[t]
\includegraphics[scale=0.3]{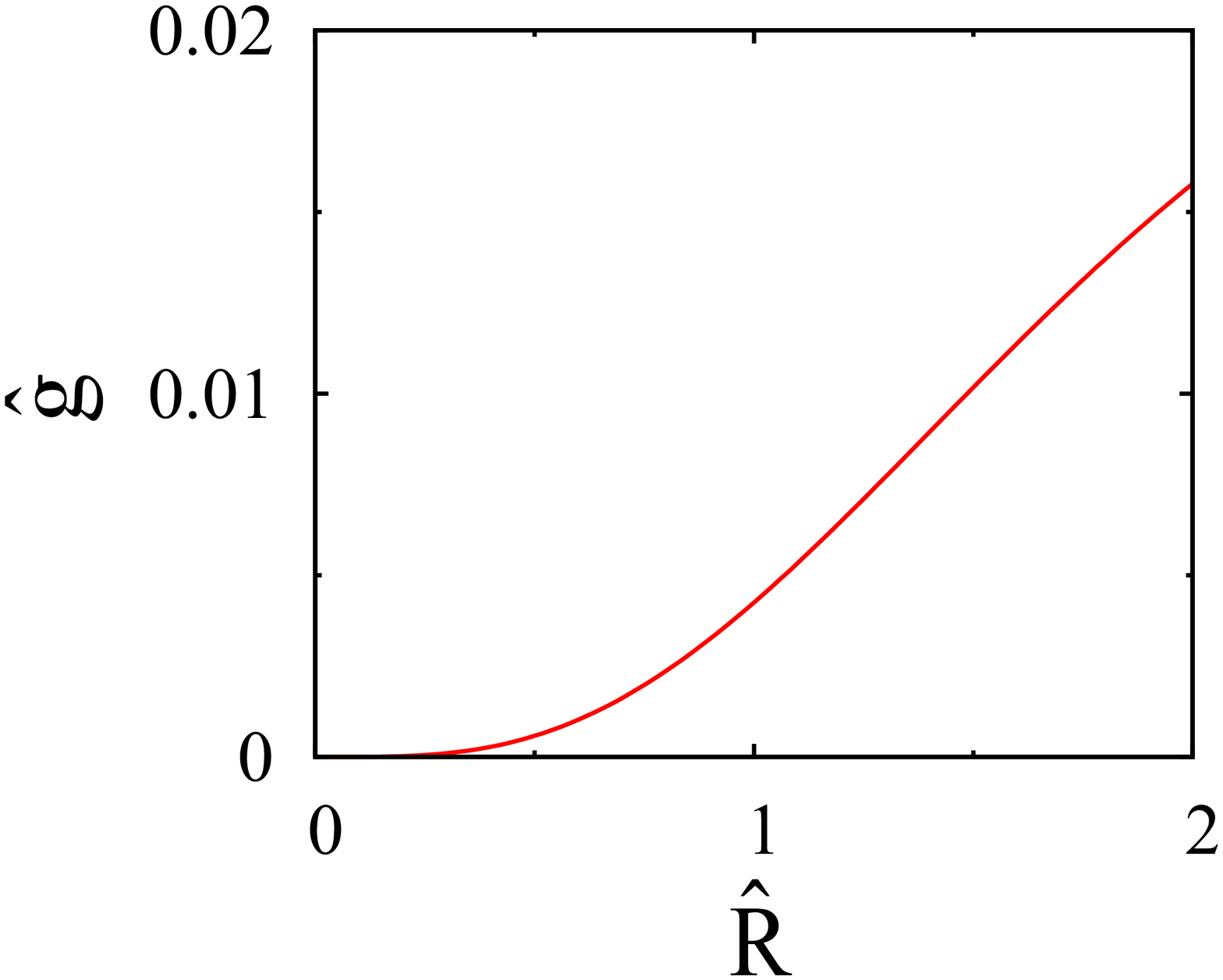}
\caption{The scaled coefficient $\hat{g}$ as a function of $\hat{R}$.}
\label{Flo:rescaled-3} 
\end{figure}

Since we have obtained the concentration profile of $c$ for a given interface configuration, we can now evaluate the 
velocities in Eqs. (\ref{u1}) and (\ref{u2}), which are carried out in Appendix \ref{the-evaluation-of}. It turns out  that there is a simple relation $\vec{u}_2=-(3/4)\vec{u}_1$. From the results obtained in Appendix \ref{the-evaluation-of}, the time-evolution equation for the center of mass is given up to the cubic non-linearity by
\begin{eqnarray}
m\dot{u}^{\alpha}=\left(-1+\tau\right)u^{\alpha}-gu^{\alpha}\left|u\right|^{2}
   \label{eqforu}
\end{eqnarray}
where
\begin{eqnarray}
  m&=&-M\frac{\partial Q_{3}}{\partial s}\Big|_{s=R},\\
   \label{m}
 \tau&=&-M\frac{\partial Q_{2}}{\partial s}\Big|_{s=R},\\
  \label{tau0}
 g&=&\frac{3M}{5}\left[-\frac{2}{R^{2}}\frac{\partial Q_{4}}{\partial s}+\frac{2}{R}\frac{\partial^{2}Q_{4}}{\partial s^{2}}+\frac{\partial^{3}Q_{4}}{\partial s^{3}}\right]_{s=R}   ,
 \end{eqnarray}
with
\begin{eqnarray}
M\equiv\frac{2\sigma_1}{15\eta_0}.
\label{M}
\end{eqnarray}
As will be shown below, all the coefficients $m$, $\tau$ and $g$ are positive.  The term proportional to $\vec{u}^{2}$ does not appear, because it is not a dissipative term. The third order term $-g\left|u\right|^{2}u^{\alpha}$
is needed to make the migration velocity finite.  By choosing $1/\beta$ as the characteristic length  and $1/(D\beta^2)$ as the characteristic time of the problem, Eq. (\ref{eqforu}) can be written in terms of the dimensionless quantities as
\begin{eqnarray}
\hat{m}\frac{d\hat{u}^{\alpha}}{d\hat{t}}=(-\tau_c+\hat{\tau})\hat{u}^{\alpha}-\hat{g}\hat{u}^{\alpha}\left|\hat{u}\right|^{2}   ,
   \label{scaledu}
\end{eqnarray}
where $\hat{t}=tD\beta^2$, $\hat{u}=u/(D\beta)$ and 
\begin{eqnarray}
\tau_c=\frac{D^2\beta^3}{MA}=\frac{15\eta_0D^2\beta^3}{2\sigma_1A}   .
   \label{tauc}
\end{eqnarray}
Here we consider the case that $ \sigma_1A$ is positive.
It is remarkable that all the parameters in the system are combined together as $\tau_c$ given by (\ref{tauc}) so that $\tau_c$ is the only dimensionless parameter. This is the case even if one takes account of the convective term in Eq. (\ref{composition}) since it does not contain any extra parameters. 
The dimensionless coefficients depend only on $\hat{R}=R\beta$ and are given by
\begin{eqnarray}
\hat{m}(\hat{R}) & =&mD\beta^2\tau_c  ,\\
\hat{\tau} (\hat{R}) & =&\tau \tau_c   ,\\
\hat{g}(\hat{R})  & =&g(D\beta)^2\tau_c  .
\end{eqnarray}
These scaled coefficients have been evaluated numerically and plotted in Figs. \ref{Flo:rescaled-1},  \ref{Flo:rescaled-2} and \ref{Flo:rescaled-3} , which indicate that those are definitely positive.

\section{Discussion}
\label{discussion}

We have formulated the theory of self-propulsion of a droplet caused by a Marangoni effect and chemical reactions.
Equation of motion for a spherical droplet has been derived as Eq. (\ref{scaledu}) which exhibits a drift bifurcation. The hydrodynamic effects are taken into consideration by the Stokes approximation for the fluid velocity. This is justified when the time variation of the concentrations is much slower than that of the local fluid velocity.  We have made two assumptions. One is the
assumption that the interface (surface of droplet) is infinitesimally thin. This assumption is satisfied when the droplet radius is much larger than the interface width. The other assumption is that the relaxation of the component $c$ is much faster than the interface motion. Since the interface velocity is arbitrarily  small in the vicinity of the drift bifurcation threshold, the second assumption is consistently justified in the theory.

\begin{figure}[t]
\includegraphics[scale=0.7]{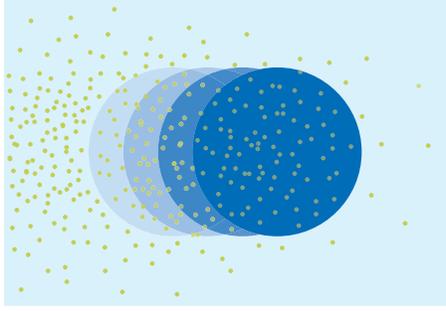}
\caption{Translational motion of a droplet. The droplet is migrating to the right under the non-uniform distribution of the $c$ component indicated by the small dots. }
\label{Flo:droplet} 
\end{figure}

The mechanism that  a droplet undergoes a translational motion in our
model for $A>0$ and $\sigma_1>0$ is as follows. When  a droplet is
motionless, there is an isotropic concentration distribution of $c$
around it. The concentration profile outside the droplet is a decreasing
function of the distance from the center of mass. Let us suppose that the position of the droplet is shifted slightly. 
Then, the concentration of $c$ decreases (increases) at the front (rear). 
If the relaxation rate of the component $c$ is infinite, this concentration unbalance is recovered instantaneously. 
However, when the relaxation is finite, the droplet tends to shift further since the interfacial energy is an increasing function of $c$. This is shown schematically in Fig. \ref{Flo:droplet}.  
In fact, it is found that the terms with the coefficients $\tau$,  $m$
and $g$ in Eq.  (\ref{eqforu}) arise from the higher order terms
($c_{\vec{q}}^{(1)}$, $c_{\vec{q}}^{(2)}$, and $c_{\vec{q}}^{(3)}$, respectively) in the short time expansion in Eq. (\ref{c}). Therefore, if the time-delayed effect $\tau u^{\alpha}$ dominates the term $-u^{\alpha}$ which corresponds to the Stokes drag force, the droplet undergoes migration.  It is noted that this argument can also be applied to the case $A<0$ and $\sigma_1<0$.

We can estimate the effect of the convective term in Eq. (\ref{composition}) which have been ignored in the treatment in section \ref{equation.motion}. In Appendix \ref{convection}, we derive the correction from the convective term up to the first order of the perturbation expansion. The coefficient $\tau$ is evaluated since this quantity is directly related to the drift instability threshold. In the limit $\hat{R} \to 0$, we obtain 
\begin{eqnarray}
\tau = \frac{1}{\tau_c } \frac{2\hat{R}^3}{15}P   .
\label{correction}
\end{eqnarray}
When the convective term is not considered, we have  $P=1$ from (\ref{tau0}). The first order correction from the convective term gives us $P=31/56$ as shown in Appendix \ref{convection}.  
Since migration of droplet occurs for $\tau \ge 1$, this indicates that the stronger Marangoni effect is necessary when the convection of the third component exists. 

The reason as to why the convective term of $H(\vec{r})\equiv\vec{v}\cdot \vec{\nabla} c$ tends to suppress the Marangoni effect can be understood as follow. Substituting the local velocity given by Eq. (\ref{localv}), we have the value of $H$ at the interface
\begin{eqnarray}
H_I = \vec{u}\cdot \vec{\nabla} c\Big|_I   ,
\label{HI}
\end{eqnarray}
When $A$ is positive, $\vec{\nabla} c_I $ and $\vec{u}$ are anti-parallel (parallel) to each other at the front (rear) of the moving droplet so that we may expect that $H<(>)0 $ at the front (rear) area. Since the first order correction to the concentration  $c$  is given by $c(\vec{r})=-[-D\nabla^2+\gamma]^{-1}H(\vec{r})$ and the operator $[-D\nabla^2+\gamma]^{-1}$ is positive definite, the concentration $c$ tends to increase (decrease) at the front (rear). This is just opposite to the concentration variation described above for the mechanism of translational motion. 

One of the characteristic features of the present theory is that all the parameters in the model equations are combined as $\tau_c$ given by
Eq. (\ref{tauc}) which determines the threshold of the drift bifurcation. Since $\tau_c$ is inversely proportional to $A$ and $\sigma_1$, the self-propulsion is easier for the stronger production of $c$ (i.e., larger values of $A$) and for stronger Marangoni effect (i.e., larger values of $\sigma_1$). Note that $\hat{\tau}$ is an increasing function of the radius of droplet. This means that the drift instability is favorable for larger droplet if other parameters are fixed and if any shape instability would not occur. 

We make a remark on the sign of the Marangoni factor. We have restricted ourselves to the case of $A\sigma_1>0$. When this quantity is negative, the coefficients $m$ and $g$ are negative in Eq. (\ref{eqforu}). Therefore, in this case,  we have to take account of the higher time derivatives and the higher nonlinear terms of $\vec{u}$. However, this is beyond our present theoretical formulation. 

In the present theory, the third component is produced inside a
droplet. However, if it is produced only on the droplet surface, the
step function in Eq. (\ref{composition}) should be replaced by the delta
function. We expect that the results  obtained in the present paper are
not essentially altered if the component  $c$ diffuses to the inside of
droplet as well as the outside. 
Such a model has been studied where the time-evolution equation of surfactant on the surface of droplet is introduced explicitly
 \cite{Yoshinaga}.

A self-propulsion of an oily droplet has
been observed in a micron size \cite{Toyota1}. In this experiment,  the
molecules which constitute the droplet are produced by a chemical
reaction which takes place at the droplet surface. 
Another experiment 
 by Thutupalli {\it et al.
\cite{Thutupalli}
} shows that an aqueous droplet of the order of 100$\mu
m$ surrounded by oil with surfactant molecules  undergoes migration by
causing a non-uniform surface tension due to bromination on its
surface. 
In these experiments, however, it seems that the bifurcation from a stationary state to a moving state predicted in the present study has not
been observed.
Further systematic experiments are desired.

Since fluid droplets are soft, they are generally deformed in migration. A coupling between migration velocity and shape deformations has been formulated recently in an excitable reaction-diffusion system \cite{SHO}. Extension of  such a theory to the present hydrodynamical system will be carried out in the future.

\section*{Acknowledgements} 
This work was supported by the JSPS Core-to-Core Program "International research network for non-equilibrium dynamics of soft matter" and  the Grant-in-Aid for the Global COE Program "The Next Generation of Physics, Spun from Universality and Emergence" from the Ministry of Education, Culture, Sports, Science and Technology (MEXT) of Japan.
TO is supported by  a Grant-in-Aid for Scientific Research (C) from Japan Society for Promotion of Science.
NY acknowledge the support by a Grant-in-Aid for Young Scientists
(B) (No.23740317).

\appendix

\section{Derivation of the forces\label{forces}}

In this Appendix, we derive the formulas  (\ref{f1}) and (\ref{f2}).
The force (\ref{f}) is written as
\begin{eqnarray}
f^{\alpha}=-\nabla^{\alpha}p-\phi\nabla^{\alpha}\frac{\delta
 F}{\delta\phi}-c\nabla^{\alpha}\frac{\delta F}{\delta c}
.
\label{force1}
\end{eqnarray}
Substituting the free energy (\ref{Free}) into Eq. (\ref{force1}), we
obtain the modified pressure 
 \begin{eqnarray}
p'=p+
\phi\frac{\partial f_{\rm GL}}{\partial \phi}
- f_{\rm GL}
+c\frac{\partial f_0}{\partial c} 
- f_0 ,
\label{p'}
\end{eqnarray}
and 
\begin{eqnarray}
f^{\alpha} 
& = & 
-\nabla^{\alpha}p'-(\nabla^{\alpha}\phi)(\nabla^{\beta}\phi)(\nabla^{\beta}B)-B(c)(\nabla^{\alpha}\phi)\nabla^{2}\phi+\frac{1}{2}(\nabla^{\alpha}B)
|\vec{\nabla}\phi|^{2}
\nonumber 
\\
 & = & 
-\nabla^{\alpha}p'+|\vec{\nabla}\phi|^{2}(\nabla^{\beta}B)(\delta_{\alpha\beta}-n^{\alpha}n^{\beta}) \nonumber \\ 
&-&B(c)(\nabla^{\alpha}\phi)\nabla^{2}\phi-\frac{1}{2}(\nabla^{\alpha}B)
|\vec{\nabla}\phi|^{2}
,
 \label{force2}
 \end{eqnarray}
where $\vec{n}=-\vec{\nabla}\phi/|\vec{\nabla}\phi|$. In the last
term on the first line of Eq. (\ref{force2}), we have used the relation
$(\nabla^{\alpha}c)\partial B/\partial c=\nabla^{\alpha}B$. Note
the formula 
\begin{eqnarray}
\nabla^{2}\phi & = & -\nabla^{\beta}\left(n^{\beta}|\vec{\nabla}\phi|\right)=-(\vec{\nabla}\cdot\vec{n})|\vec{\nabla}\phi|+n^{\beta}\nabla^{\beta}|\vec{\nabla}\phi|\nonumber \\
 & = &  -(\vec{\nabla}\cdot\vec{n})|\vec{\nabla}\phi|+n^{\gamma}\nabla^{\gamma}(n^{\beta}\nabla^{\beta}\phi) \nonumber \\
  & = &  -(\vec{\nabla}\cdot\vec{n})|\vec{\nabla}\phi|+n^{\gamma}n^{\beta}\nabla^{\gamma}(\nabla^{\beta}\phi)+n^{\gamma}|\nabla\phi|n^{\beta}\nabla^{\gamma}(n^{\beta}) \nonumber \\
 & = & -(\vec{\nabla}\cdot\vec{n})|\vec{\nabla}\phi|+n^{\gamma}n^{\beta}(\nabla^{\gamma}\nabla^{\beta}\phi)  ,
 \label{curvature}
 \end{eqnarray}
where we have used the fact that $n^{\beta}(\nabla^{\gamma}n^{\beta})=(1/2)\nabla^{\gamma}(n^{\beta})^{2}=0$
since $(n^{\beta})^{2}=1$. 
 Substituting this into Eq. (\ref{force2}), we obtain 
 \begin{eqnarray}
f^{\alpha} & = & -\nabla^{\alpha}p'+|\vec{\nabla}\phi|^{2}(\nabla^{\beta}B)(\delta_{\alpha\beta}-n^{\alpha}n^{\beta})-B(c)n^{\alpha}(\vec{\nabla}\cdot\vec{n})|\vec{\nabla}\phi|^{2}    \nonumber \\
 & - & \frac{1}{2}(\nabla^{\alpha}B)(\vec{\nabla}\phi)^{2}+B(c)n^{\alpha}|\vec{\nabla}\phi|n^{\gamma}n^{\beta}(\nabla^{\gamma}\nabla^{\beta}\phi) \nonumber \\
 &+&\frac{1}{2}\nabla^{\alpha}\left(B(\vec{\nabla}\phi)^{2}\right) -\frac{1}{2}\nabla^{\alpha}\left(B(\vec{\nabla}\phi)^{2}\right)\nonumber \\
 & = & -\nabla^{\alpha}p''+|\vec{\nabla}\phi|^{2}(\nabla^{\beta}B)(\delta_{\alpha\beta}-n^{\alpha}n^{\beta})-B(c)n^{\alpha}(\vec{\nabla}\cdot\vec{n})|\vec{\nabla}\phi|^{2}\nonumber \\
 & - & B(c)n^{\alpha}|\vec{\nabla}\phi|n^{\gamma}n^{\beta}(\nabla^{\gamma}\nabla^{\beta}\phi)+B(c)(\nabla^{\alpha}\nabla^{\beta}\phi)(\nabla^{\beta}\phi)\nonumber \\
 & = & -\nabla^{\alpha}p''+|\vec{\nabla}\phi|^{2}(\nabla^{\beta}B)(\delta_{\alpha\beta}-n^{\alpha}n^{\beta})-B(c)n^{\alpha}(\vec{\nabla}\cdot\vec{n})|\vec{\nabla}\phi|^{2}\nonumber \\
 & + & B(c)(\nabla^{\gamma}\nabla^{\beta}\phi)(\nabla^{\beta}\phi)(\delta_{\alpha\gamma}-n^{\alpha}n^{\gamma})  ,
 \label{force3}
 \end{eqnarray}
 where 
 \begin{eqnarray}
p''=p'+\frac{1}{2}B(\nabla\phi)^{2}  .
\label{p''}
\end{eqnarray}
 Therefore the force $\vec{f}$ can be divided into the normal and
the perpendicular components 
\begin{eqnarray}
f^{\alpha} & = & -\nabla^{\alpha}p''+f_{\parallel}^{\alpha}+f_{\perp}^{\alpha}  ,
\end{eqnarray}
 where 
 \begin{eqnarray}
f_{\parallel}^{\alpha}=-n^{\alpha}B(c)|\vec{\nabla}\phi|^{2}(\vec{\nabla}\cdot\vec{n})  ,
\label{force4}
\end{eqnarray}
 \begin{eqnarray}
f_{\perp}^{\alpha}=(\delta_{\alpha\beta}-n^{\alpha}n^{\beta})\Big[(\nabla^{\beta}B)|\vec{\nabla}\phi|^{2}-B(c)|\vec{\nabla}\phi|(\nabla^{\beta}\nabla^{\gamma}\phi)n^{\gamma}\Big]  .
\label{force5}
\end{eqnarray}
 The second term in Eq. (\ref{force5}) is negligible compared to
the first term in the sharp interface limit. In fact, we have 
\begin{eqnarray}
|\vec{\nabla}\phi|(\nabla^{\beta}\nabla^{\gamma}\phi)n^{\gamma}]&=&|\vec{\nabla}\phi|\left[(\nabla^{\beta}n^{\gamma})|\vec{\nabla}\phi| 
+n^{\gamma}(\nabla^{\beta}|\vec{\nabla}\phi|)\right]n^{\gamma} \nonumber \\
&=&\frac{1}{2}\nabla^{\beta}|\vec{\nabla}\phi|^{2}  ,
\label{force6}
\end{eqnarray}
 where we have again used the formula $n^{\gamma}(\nabla^{\beta}n^{\gamma})=(1/2)\nabla^{\beta}(n^{\gamma})^{2}=0$. The integral of $B(c)\nabla^{\beta}|\vec{\nabla}\phi|^{2}$
across the interface vanishes provided that $B$ varies weakly across
the interface. Therefore we ignore the second term in Eq. (\ref{force5}).

\section{Derivation of the migration velocity \label{Derivation-of-velocity}}

In this Appendix, we derive Eqs.  (\ref{u1}) and (\ref{u2}).
In order to obtain Eq.  (\ref{u1}), the following formula for a spherical droplet \cite{Ohta1} is necessary.
\begin{eqnarray}
\int da'n^{\alpha}(a)T^{\alpha\beta}\left(\vec{r}(a),\vec{r}(a')\right)n^{\beta}(a')Y_{lm}(a')=E_{l}Y_{lm}(a)  ,
\end{eqnarray}
 where
  \begin{eqnarray}
E_{l}=\frac{R}{\eta_{0}}\frac{2l(l+1)}{(2l-1)(2l+1)(2l+3)},
\end{eqnarray}
 and $Y_{lm}(a')$ is the spherical harmonics. The representation of the unit
vector $\vec{n}$  in terms of $Y_{1,m}$ is also necessary.
\begin{eqnarray}
\vec{n} & = & \left(\sin\theta\cos\phi,\sin\theta\sin\phi,\cos\theta\right)\\
 & = & \left(\sqrt{\frac{2\pi}{3}}\left(-Y_{11}+Y_{1-1}\right),i\sqrt{\frac{2\pi}{3}}\left(Y_{11}+Y_{1-1}\right),\sqrt{\frac{4\pi}{3}}Y_{10}\right)  .\end{eqnarray}
Applying these formulas to Eq. (\ref{upara}), one can carry out the integral over $a$ so that Eq.  (\ref{u1}) is obtained.

Next we calculate Eq. (\ref{uperp}). First we make an ansatz as
\begin{eqnarray}
\int da'T^{\alpha\beta}\left(\vec{r}(a),\vec{r}(a')\right)n^{\beta}(a')n^{\gamma}(a')=X\delta_{\alpha\gamma}
+Yn^{\alpha}(a)n^{\gamma}(a)  .
\label{perpformula}
\end{eqnarray}
The unknown constants $X$ and $Y$ are determined as follows. We note the identities;
 \begin{eqnarray}
\int da'T^{\alpha\beta}\left(\vec{r}(a),\vec{r}(a')\right)n^{\beta}(a')n^{\alpha}(a')=3X+Y ,  \\
\int da'n^{\alpha}(a)T^{\alpha\beta}\left(\vec{r}(a),\vec{r}(a')\right)n^{\beta}(a')n^{\gamma}(a)n^{\gamma}(a')=X+Y  .
\end{eqnarray}
The left hand side of these expressions is readily evaluated as
\begin{eqnarray}
& &\int da'T^{\alpha\beta}\left(\vec{r}(a),\vec{r}(a')\right)n^{\beta}(a')n^{\alpha}(a')  \nonumber \\
&=& \frac{R}{8\eta_0}\int_{-1}^{1}d(\cos\theta)\frac{1+\sin^{2}(\theta/2)}{\sin(\theta/2)}  =\frac{2R}{3\eta_0}  ,  
 \end{eqnarray}
 \begin{eqnarray}
& &\int da'n^{\alpha}(a)T^{\alpha\beta}\left(\vec{r}(a),\vec{r}(a')\right)n^{\beta}(a')n^{\gamma}(a)n^{\gamma}(a') \nonumber \\
&=& \frac{R}{8\eta_0}\int_{-1}^{1}d(\cos\theta)\cos\theta\frac{\cos\theta-\sin^{2}(\theta/2)}{\sin(\theta/2)} =\frac{4R}{15\eta_0}  ,
 \end{eqnarray}
 where $\theta(>0)$ is the angle between $\vec{n}(a)$ and $\vec{n}(a')$.
Therefore we obtain
 \begin{eqnarray} 
X=\frac{R}{5\eta_0}  ,
\label{X}\\
Y=\frac{R}{15\eta_0}   .
\label{Y}
 \end{eqnarray}
 By using the formula (\ref{perpformula}), Eq. (\ref{u2}) is readily obtained.

\section{Derivation of the coefficients \label{the-evaluation-of}}

In this section, we derive the migration velocities by  evaluating Eqs. (\ref{u1}) and (\ref{u2}).
Substituting Eqs.  (\ref{c0}), (\ref{c1}), (\ref{c2}) and (\ref{c3})
into Eq. (\ref{u1}), we obtain
\begin{eqnarray}
u_1^{\alpha}=u_1^{(1){\alpha}}+u_1^{(2){\alpha}}+u_1^{(3){\alpha}}  ,
 \end{eqnarray}
where
\begin{eqnarray}
u_{1}^{(1)\alpha} & = & -\frac{2E_{1}\sigma_{1}}{\Omega }u^{\beta}\frac{\partial Q_{2}(s)}{\partial s}\Big|_{s=R}\int da'n^{\alpha}n^{\beta}\nonumber \\ 
 & = & -\frac{2E_{1}\sigma_{1}}{R}u^{\alpha}\frac{\partial Q_{2}(s)}{\partial s}\Big|_{s=R}  ,
 \label{u11}
 \end{eqnarray}
 \begin{eqnarray}
u_{1}^{(2)\alpha} & = & \frac{2E_{1}\sigma_{1}}{\Omega}\dot{u}^{\beta}\frac{\partial Q_{3}(s)}{\partial s}\Big|_{s=R}\int da'n^{\alpha}n^{\beta} \nonumber \\
 & = & \frac{2E_{1}\sigma_{1}}{R}\dot{u}^{\alpha}\frac{\partial Q_{3}(s)}{\partial s}\Big|_{s=R}  ,
 \label{u12}
  \end{eqnarray}
   \begin{eqnarray}
u_{1}^{(3)\alpha} & = & -\frac{2E_{1}\sigma_{1}}{\Omega}u^{\delta}u^{\beta}u^{\gamma}\int da'n^{\alpha}\frac{\partial}{\partial s^{\delta}}\frac{\partial}{\partial s^{\beta}}\frac{\partial}{\partial s^{\gamma}}Q_{4}(s) \nonumber \\
 & = & -\frac{2E_{1}\sigma_{1}}{R}\left[-\frac{6}{5R^{2}}\frac{\partial Q_{4}}{\partial s}+\frac{6}{5R}\frac{\partial^{2}Q_{4}}{\partial s^{2}}+\frac{3}{5}\frac{\partial^{3}Q_{4}}{\partial s^{3}}\right]u^{\alpha}\left|u\right|^{2}  ,
 \label{u13}
 \end{eqnarray}
 with $E_1=4R/(15\eta_0)$.
In these derivations, we have used the following relations;
\begin{eqnarray}
\frac{R}{\Omega}\int dan^{\alpha}n^{\beta}n^{\gamma}n^{\delta} & = & \frac{1}{5}\left(\delta_{\alpha\beta}\delta_{\gamma\delta}+\delta_{\alpha\gamma}\delta_{\beta\delta}+\delta_{\alpha\delta}\delta_{\beta\gamma}\right)   ,   \\
\frac{R}{\Omega}\int dan^{\alpha}n^{\beta} & = & \delta_{\alpha\beta}  .
\end{eqnarray}

In order to calculate $u_2^{\alpha}$ in Eq. (\ref{u2}), we need the gradient of the concentration $c$.
\begin{eqnarray}
\nabla^{\gamma}c^{(1)} & = & u^{\alpha}\frac{\partial}{\partial s^{\gamma}}\frac{\partial}{\partial s^{\alpha}}Q_{2}(s) \nonumber \\
&=&u^{\alpha}\left[\frac{1}{R}\left(\delta_{\gamma\alpha}-n^{\alpha}n^{\gamma}\right)\frac{\partial Q_{2}}{\partial s}+n^{\alpha}n^{\gamma}\frac{\partial^{2}Q_{2}}{\partial s^{2}}\right]   ,   \\
\nabla^{\gamma}c^{(2)} & = & -\dot{u}^{\alpha}\frac{\partial}{\partial s^{\gamma}}\frac{\partial}{\partial s^{\alpha}}Q_{3}(s)+u^{\alpha}u^{\beta}\frac{\partial}{\partial s^{\gamma}}\frac{\partial}{\partial s^{\alpha}}\frac{\partial}{\partial s^{\beta}}Q_{3}(s) \nonumber \\
 & = & -\dot{u}^{\alpha}\left[\frac{1}{R}\left(\delta_{\gamma\alpha}-n^{\alpha}n^{\gamma}\right)\frac{\partial Q_{3}}{\partial s}+n^{\alpha}n^{\gamma}\frac{\partial^{2}Q_{3}}{\partial s^{2}}\right]  ,  \\
\nabla^{\gamma}c^{(3)} & = & u^{\alpha}u^{\beta}u^{\delta}\frac{\partial}{\partial s^{\gamma}}\frac{\partial}{\partial s^{\alpha}}\frac{\partial}{\partial s^{\beta}}\frac{\partial}{\partial s^{\delta}}Q_{4}(s)     \nonumber \\
 & = & +\frac{3}{R^{3}}\Big[-u^{\gamma}\left|u\right|^{2}+3\left|u\right|^{2}u^{\alpha}n^{\alpha}n^{\gamma} \nonumber \\
 &+& 3u^{\gamma}u^{\alpha}u^{\beta}n^{\alpha}n^{\beta}-5u^{\alpha}u^{\beta}u^{\delta}n^{\alpha}n^{\beta}n^{\delta}n^{\gamma}\Big]\Big(\frac{\partial Q_{4}}{\partial s}-R\frac{\partial^{2}Q_{4}}{\partial s^{2}} \Big) \nonumber \\
 &  +& \frac{3}{R}\left[\left|u\right|^{2}u^{\alpha}n^{\alpha}n^{\gamma}+u^{\gamma}u^{\alpha}u^{\beta}n^{\alpha}n^{\beta}-2u^{\alpha}u^{\beta}u^{\delta}n^{\alpha}n^{\beta}n^{\gamma}n^{\delta}\right]\frac{\partial^{3}Q_{4}}{\partial s^{3}} \nonumber \\
 & + & u^{\alpha}u^{\beta}u^{\delta}n^{\gamma}n^{\alpha}n^{\beta}n^{\delta}\frac{\partial^{4}Q_{4}}{\partial s^{4}}  .
 \end{eqnarray}
Substituting these into Eq. (\ref{u2}), we obtain
\begin{eqnarray}
u_{2}^{(1)\alpha} &= &\frac{2 X\sigma_{1}}{R}u^{\alpha}\frac{\partial Q_{2}}{\partial s}\Big|_{s=R}  ,
\label{u21}\\
u_{2}^{(2)\alpha}&=&-\frac{2X\sigma_{1}}{R}\dot{u}^{\alpha}\frac{\partial
Q_{3}}{\partial s}\Big|_{s=R}  ,
\label{u22}\\
u_{2}^{(3)\alpha} & = &  \frac{2X\sigma_{1}}{R}\left[-\frac{1}{R^{2}}\frac{6}{5}\frac{\partial Q_{4}}{\partial s}+\frac{1}{R}\frac{6}{5}\frac{\partial^{2}Q_{4}}{\partial s^{2}}+\frac{3}{5}\frac{\partial^{3}Q_{4}}{\partial s^{3}}\right]_{s=R}\left|u\right|^{2}u^{\alpha}  ,
\label{u23}
 \end{eqnarray}
where $X$ has been defined by Eq. (\ref{X}).  Comparing Eqs. (\ref{u11})-(\ref{u13}) with Eqs. (\ref{u21})-(\ref{u23}), we note that
$\vec{u}_2=-(3/4)\vec{u}_1$.

\section{Correction from the convective term\label{convection}}
In this Appendix, we calculate the coefficient $\tau$ by taking account of the correction from the convective term in Eq. (\ref{composition}).
Up to the first order of $\vec \nabla \cdot (\vec{v}c)$, Eq. (\ref{tau0}) has an additive correction as
\begin{eqnarray}
\tau u^{\alpha} & = & -M\frac{\partial Q_{2}}{\partial s}\Big|_{s=R}u^{\alpha} \nonumber \\
 & -& \frac{MR}{\Omega}\int da'n^{\alpha}\left[\int_{\vec{q}}G_{q}e^{-i\vec{q}\cdot\left(\vec{r}_{G}+\vec{s}\right)}\int_{\vec{r'}}d^{3}re^{i\vec{q}\cdot\vec{r'}}\left(\vec{v}(\vec{r'})\cdot i\vec{q}c^{\left(0\right)}(\vec{r'})\right)\right]  ,
 \label{tau1}
 \end{eqnarray}
where we have used the relation $\vec{u}_2=-(3/4)\vec{u}_1$. The vector $\vec{v}(\vec{r})$ in the second term is the velocity field around (and inside) the droplet moving at a constant velocity $u$ along the $z$-axis and is given by  \cite{Onuki}
\begin{align}
\vec{v}(\vec{r}_{G}+\vec{r}) 
& = 
\begin{cases}
u\left[\left(\frac{5}{2}-\frac{3r^{2}}{R^{2}}\right)\mathbf{e}_{z}+\frac{3z}{2R^{2}}\mathbf{r}\right]\:(for~r<R)
\\
 u\left[-\left(\frac{R^{3}}{2r^{3}}\right)\mathbf{e}_{z}+\frac{3R^{3}z}{2r^{5}}\mathbf{r}\right]\:\left(for~r>R\right).
\end{cases}
\label{localv}
 \end{align}
Analytical evaluation of  the integrals in Eq. (\ref{tau1}) seems impossible in a general condition. Here we consider the limit $\hat{R}=\beta R \to 0$. In this case, we may approximate $G(\vec{r})$ as  $G(\vec{r})=1/(4\pi D r)$ and $\tau$ is calculated as 
\begin{eqnarray}
\tau&=& -\frac{MR}{\Omega}\frac{4\pi}{27D^{2}}AR^{5}\left[\frac{3}{2}\left(\frac{1}{5}-\frac{1}{7}\right)-\left(1-\frac{1}{4}\right)\right] \nonumber \\
 & = & \frac{MR}{\Omega}\frac{4\pi}{27D^{2}}R^{5}A\frac{93}{140}  .
 \end{eqnarray}
If the second term in Eq. (\ref{tau1}) is ignored, the factor 93/140 is replaced by 6/5.

We can also calculate the coefficient $m$ by taking account of the correction from the convective term in Eq. (\ref{composition}). 
\begin{eqnarray}
&m&\dot{u}^{\alpha} = -M\frac{\partial Q_{3}}{\partial s}\Big|_{s=R}\dot{u}^{\alpha} \nonumber \\
 &+  & \frac{MR}{\Omega}\int da'n^{\alpha}\left[-\int_{\vec{q}}G_{q}^{2}e^{-i\vec{q}\cdot\left(\vec{r}_{G}+\vec{s}\right)}\frac{\partial}{\partial t}\int d^{3}re^{i\vec{q}\cdot\vec{r'}}\left(\vec{v}(\vec{r'})\cdot i\vec{q}c^{\left(0\right)}(\vec{r'})\right)\right]  .
 \end{eqnarray}
The lowest order contribution from the first term is given by
\begin{equation}
-M\frac{\partial}{\partial s}Q_{3}(s)\Big|_{s=R}=\Big[\frac{AMR^{5}}{24 D^{3}}\frac{1}{\hat{R}}+\left(terms~ finite~ for~ \hat{R}\rightarrow0\right)\Big].
\end{equation}
The second term due to the convection of the composition $c$ has no term which is infinite for $\beta \to 0$.
Thus, the contribution to the coefficient $m$ from the convection of component $c$  is found to be higher order of $\beta R$. We expect the same situation for $g$ but have not confirmed it since the expression is very complicated.
Finally, we make a remark  that the smallness of  $\varepsilon$ in Eq. (\ref{EOMofC}) is independent of the smallness of $\hat{R}$.


\end{document}